\documentclass[12pt]{article}
\usepackage{geometry}
\usepackage{enumitem,amssymb}
\usepackage{ragged2e}
\usepackage{pifont}
\newcommand{\cmark}{\ding{51}}%
\newcommand{\done}{\rlap{$\square$}{\raisebox{2pt}{\large\hspace{1pt}\cmark}}%
\hspace{-2.5pt}}

\newlist{thematic}{itemize}{8}
\setlist[thematic]{label=\done}

\usepackage{booktabs} 
\usepackage{array} 
\usepackage{verbatim} 
\usepackage{graphicx}
\usepackage[colorlinks,
            linkcolor=blue,
            citecolor=blue,
            filecolor=blue,
            urlcolor=blue]{hyperref} 
\usepackage{tabularx} 
\usepackage[table]{xcolor} 
\usepackage{wasysym} 
\usepackage{subfigure} 
\usepackage{enumitem}
\usepackage{lineno}

\usepackage[bibstyle = numeric-comp, sortcites = true, natbib, sorting=none, defernumbers=false, uniquelist=false, url=false, eprint=false, isbn=false, date=year, giveninits=true, backend=bibtex]{biblatex}

\usepackage{setspace}
\usepackage{cleveref}

\begin{document}
\thispagestyle{empty}
\large
\noindent \textbf{Astro2020 Science White Paper} 
\LARGE
\vspace{-0.1cm}
\begin{center}
\noindent \textbf{Pulsars in a Bubble?} \\
\noindent \textbf{Following Electron Diffusion in the Galaxy\\ with TeV Gamma Rays} \par
\end{center}
\normalsize
\vspace{-0.1cm}
\noindent \textbf{Thematic Areas:} 
\vspace{-0.2cm}
\begin{thematic}[noitemsep]
\item Formation and Evolution of Compact Objects
\item Cosmology and Fundamental Physics
\item Multi-Messenger Astronomy and Astrophysics
\end{thematic} 
\setcounter{page}{0}
\vspace{0.2cm}

\noindent \textbf{Principal Author:}

\noindent Name: H. Fleischhack
					
\noindent Institution: Michigan Technological University (MTU)

\noindent Email: \href{mailto:hfleisch@mtu.edu}{hfleisch@mtu.edu}\\
\noindent Phone: 906 487 2124\\

\vspace{-0.3cm}
\noindent\textbf{Co-authors/Endorsers:}
\href{mailto:amalbert@lanl.gov}{A. Albert} (Los Alamos National Lab [LANL]),
\href{mailto:cesar.alvarez@unach.mx}{C. Alvarez} (UNACH, M\'exico), 
\href{mailto:roberto.arceo@unach.mx}{R. Arceo} (UNACH, M\'exico),
\href{mailto:hgayala@psu.edu}{H. A. Ayala Solares} (Pennsylvania State University [PSU]),
\href{mailto:beacom.7@osu.edu}{J.~F.~Beacom} (The Ohio State University [OSU]),
\href{mailto:ralphbird@astro.ucla.edu}{R.~Bird} (University of California, Los Angeles), 
\href{mailto:cbrisboi@mtu.edu}{C.~A.~Brisbois} (MTU), 
\href{mailto:karen.scm@gmail.com}{K.~S.~Ca\-ballero-Mora} (UNACH, M\'exico),
\href{mailto:alberto@inaoep.mx}{A. Carrami\~nana} (INAOE, M\'exico),
\href{mailto:sabrina.casanova@mpi-hd.mpg.de}{S. Casanova} (Institute of Nuclear Physics Polish Academy of Sciences [IFJ]),
\href{mailto:pierre.cristofari@gssi.it}{P. Cristofari} (Gran Sasso Science Institute),
\href{mailto:paolo.coppi@yale.edu}{P. Coppi} (Yale University),
\href{mailto:dingus@lanl.gov}{B. L. Dingus} (LANL), 
\href{mailto:duvernois@icecube.wisc.edu}{M. A. DuVernois} (University of Wisconsin), 
\href{mailto:klengel@umd.edu}{K. L. Engel} (University of Maryland [UMD]),
\href{mailto:goodman@umd.edu}{J. A. Goodman} (UMD),
\href{mailto:green@liv.ac.uk}{T. Greenshaw} (University of Liverpool, UK), 
\href{mailto:jpharding@lanl.gov}{J.~P.~Harding} (LANL), 
\href{mailto:bhona@mtu.edu}{B. Hona} (MTU),
\href{mailto:petra@mtu.edu}{P. H. Huentemeyer} (MTU), 
\href{mailto:hli@lanl.gov}{H. Li} (LANL),
\href{mailto:linden.70@osu.edu}{T. Linden} (OSU),
\href{mailto:gilgamesh@upp.edu.mx}{G. Luis-Raya} (Universidad Polit\'ecnica de Pachuca, Mexico),
\href{mailto:malone@psu.edu}{K. Malone} (LANL),
\href{mailto:macj@cic.ipn.mx}{J. Mart\'inez-Castro} (Centro de Investigaci\'on en Computaci\'on-IPN, M\'exico),
\href{mailto:miguel@psu.edu}{M.~A.~Mostaf\'a} (PSU),
\href{mailto:mnisa@ur.rochester.edu}{M.~U.~Nisa} (University of Rochester),
\href{mailto:riviere@umd.edu}{C.~Rivi\`{e}re} (UMD),
\href{mailto:francisco.salesa@ifj.edu.pl}{F. Salesa Greus} (IFJ),
\href{mailto:asandoval@fisica.unam.mx}{A. Sandoval} (Instituto de F\'isica, UNAM, M\'exico),
\href{mailto:asmith8@umd.edu}{A. J. Smith} (UMD),
\href{mailto:Wayne.Springer@m.cc.utah.edu}{W. Springer} (University of Utah),
\href{mailto:sudoh@astron.s.u-tokyo.ac.jp}{T. Sudoh} (University of Tokyo),
\href{mailto:tollefson@pa.msu.edu}{K. Tollefson} (Michigan State University),
\href{mailto:zepeda@fis.cinvestav.mx}{A. Zepeda} (Cinvestav),
\href{mailto:hao@lanl.gov}{H. Zhou} (LANL)\\
\vspace{-0.3cm}
~\\
\noindent \textbf{Abstract:} TeV Halos, extended regions of TeV gamma-ray emission around middle-aged pulsars, have recently been established as a new source class in gamma-ray astronomy. These halos have been attributed to relativistic electrons and positrons that have left the acceleration region close to the pulsar and are diffusing in the surrounding medium.

Measuring the morphology of TeV Halos enables for the first time a direct measurement of the electron diffusion on scales of tens of parsecs. There are hints that the presence of relativistic particles affects the diffusion rate in the pulsars' surroundings. Understanding electron diffusion is necessary to constrain the origins of the apparent ``excess'' of cosmic-ray positrons at tens of GeV. TeV Halos can also be used to find mis-aligned pulsars, as well as study certain properties of the Galaxy's pulsar population.

Future VHE gamma-ray instruments will detect more of those TeV Halos and determine how much pulsars contribute to the observed cosmic-ray electron and positron fluxes, and how they affect diffusion in their environments.

\normalsize

\clearpage

\setlength\itemsep{-0.1cm}

\section{Science}

\subsection{TeV Halos}
The HAWC observatory has discovered `halos' of extended (several degrees) TeV gamma-ray emission around the Geminga and Monogem pulsars \cite{Abeysekara2017d}. For Geminga, this emission is orders of magnitudes more extended than what is expected from the PWN.

These objects, dubbed `TeV Halos', constitute a new source class \cite{Linden2017}. The emission is consistent with inverse Compton (IC) emission from electrons and positrons\footnote{In the following, collectively referred to as `electrons'.} that were originally accelerated by the pulsar or in the pulsar wind nebula (PWN), but have since escaped the acceleration region and are diffusing in the surrounding medium.  
In the region around Geminga, the diffusion radius of electrons with energies below 1 TeV is limited by the pulsar age; for higher energies, it is limited by the electron cooling time. The largest diffusion radius of about 70 pc is reached by 1 TeV electrons \cite{Abeysekara2017d}.

HAWC was able to measure the diffusion coefficient inside the Geminga and Monogem Halos. The results were significantly lower than what expected for the interstellar medium (ISM) \cite{Abeysekara2017d}. Assuming the diffusion coefficient is constant over a larger region, these measurements indicate that Geminga and Monogem can not be the primary sources of the apparent excess in cosmic-ray positrons (see \cref{sec:positrons}), as the electrons accelerated by these pulsars would cool long before reaching Earth. 

However, there are models by which the streaming of the relativistic electrons itself suppresses the diffusion coefficient near the pulsars \cite{Evoli2018}. There could be `bubbles' of suppressed diffusion extending for tens of parsecs around middle-aged pulsars, with increased diffusion in the rest of the Galaxy. In that case, electrons leaving this bubble could still propagate to Earth and significantly contribute to the local positron fraction \cite{Hooper2018}.

It has been proposed that all middle-aged pulsars should have such halos with approximately the same efficiency to produce TeV gamma rays and the same electron diffusion coefficient \cite{Linden2017}. The TeV gamma-ray flux would then only depend on the pulsar's spindown luminosity and its distance.
In that case, many more such objects should be discovered in future searches \cite{Linden2017}. Since the discovery of the first TeV Halos, HAWC has already reported at least two more TeV Halo candidates \cite{2017ATel10941....1R, 2018ATel12013....1B}. Some of the unidentified or PWN-associated sources seen by HAWC and H.E.S.S. in the galactic plane may also be TeV Halos \cite{Linden2017}.

The presence of TeV Halos may be the best observational signature for the discovery mis-aligned pulsars, whose radio beam does not strike Earth \cite{Linden2017}. The total number of TeV Halos can provide constraints on some properties of the Galactic pulsar population and the time it takes for a TeV Halo to form \cite{Sudoh:2019lav}, which is not yet known. An upper limit for the halo formation time is given by the characteristic age of the Monogem pulsar (110 kyr). 

Unresolved TeV Halos would contribute to the diffuse gamma-ray background. It has been proposed that they indeed make up a significant fraction of the measured Galactic diffuse flux at TeV energies \cite{Linden:2017blp}. That could explain the apparent discrepancies between the measured TeV diffuse flux and extrapolations from measurements at lower energies \cite{Abdo:2008if, TeVExcess}.


\subsection{Particle acceleration in pulsars and their environments}
\label{sec:pulsars}
Pulsars are rapidly rotating neutron stars with strong magnetic fields capable of particle acceleration \cite{GoldreichJulian}. They can produce pairs of electrons and positrons and accelerate them up to TeV energies via polar cap cascades \cite{Harding2018}. These particles escape the light cylinder and propagate in the surrounding medium along open field lines. Many young and middle-aged pulsars are surrounded by pulsar wind nebulae, parsec-scale regions energetically dominated by relativistic $e^+e^-$ winds from the pulsar. Once in the PWN, electrons can be further accelerated to higher energies. They emit X-rays via synchrotron emission and gamma rays via IC scattering \cite{Gaensler2006}. Existing instruments have seen gamma rays at tens of TeV from such systems (e.g., \cite{Aharonian2006_HESSJ1826,Abeysekara2017d,Abeysekara2018}). In at least one system it is possible to observe the escape of electrons from the PWN seen in X-rays \cite{Aharonian2005} and their subsequent cooling as they travel further from the pulsar \cite{Aharonian2006_HESSJ1826}. More recent work has suggested pulsars may inject particles as high as $\sim30$ TeV into its PWN \cite{Harding2018}, although there is evidence to suggest electron energies in the PWN reach at least as high as 100 TeV \cite{Abeysekara2018,Aharonian2006_HESSJ1826,Yuksel2009}. 


\subsection{Cosmic-ray electrons and the positron excess}
\label{sec:positrons}
While the bulk of cosmic rays constantly bombarding Earth are protons and nuclei, they also contain electrons (and positrons). Cosmic-ray electrons (CREs) have been observed with energies up to tens of TeV by space- and ground-based detectors. Their energies follow a soft, broken-powerlaw spectrum, with the index softening from -3 to -3.8 at around 1 TeV \cite{2017Natur.552...63D, PhysRevLett.120.261102, HESS_electrons}. At TeV to PeV energies, electrons propagating through the Galaxy cool very efficiently via synchrotron emission and IC scattering. Thus, \emph{primary} CREs at these energies must be produced in nearby sources (within hundreds of parsecs). There is also a component of \emph{secondary} electrons, produced in interactions of hadronic cosmic rays with the interstellar medium (ISM). 

At TeV energies, cosmic-ray positrons are expected to consist mainly of secondary particles, with a softer spectrum than primary electrons, and the positron fraction is expected to decrease with energy. However, several experiments have observed an increase in the positron fraction above about 10 GeV, continuing up to at least 100 GeV \cite{Adriani:2008zr,2012PhRvL.108a1103A,Accardo:2014lma}, consistent with a hardening in the positron spectrum \cite{PhysRevLett.122.041102}. 
Possible explanations comprise physics beyond the standard model, for example dark matter decay or annihilation processes (see e.g., Ref.\,\cite{Pospelov:2008jd}), as well as the presence of one or several nearby positron accelerators (see e.g., Ref.\,\cite{Hooper2017}). 
As described in \cref{sec:pulsars}, pulsars and their environments are good candidates for such accelerators. 
The study of particle acceleration in pulsars and their environment will provide important clues about the origin of the positron excess.

CREs are deflected by Galactic magnetic fields and thus do not point back to their sources. However, their sources can be identified by the electromagnetic emission produced by CREs interacting  with matter or radiation fields in and near their sources.  TeV-energy electrons in particular can produce VHE (very-high-energy, E$>$100 GeV) gamma rays via IC upscattering of lower-energy photons \cite{hintonreview}.


\clearpage 
\section{Future Research}
\label{sec:studies}
TeV Halos have only recently been described as a source class, and not much is known about them. In this section, we propose three key projects, to be conducted over the next decade or two, that will help answer the following questions:
\begin{enumerate}
    \item How do TeV Halos form and evolve?
    \item How do they affect electron diffusion in the Galaxy?
    \item What can we learn about the Galactic pulsar population from studying TeV Halos?
\end{enumerate}
The TeV gamma-ray channel is ideally suited for the study of TeV Halos. Their VHE IC emission is dominated by interactions with the CMB, a photon field of known magnitude and spectrum. Thus, there is little uncertainty in inferring the properties of the electron population from the measured IC emission. 

The electrons in the Halo can also emit synchrotron radiation in the hard X-ray/soft gamma-ray range. Detection of this emission could also lead to important physics results as it is sensitive to the magnitude of the magnetic field in the halo region.

\subsection{Deep observations of well-resolved TeV Halos}

Further studies of bright, well-resolved TeV Halos, such as the prototypical Geminga Halo, are needed. Future observations should cover a larger energy range (tens of GeV to hundreds of TeV) with high sensitivity, have improved angular and energy resolution, and be able to address the following questions:
\begin{itemize}[noitemsep]
    \item Does electron diffusion proceed homogeneously in the pulsar environment, or are there any signs for asymmetries/inhomogeneities?
    \item Is there an `edge' to the bubble, beyond which diffusion coefficient increases to the expected levels from the ISM? How far does the bubble extend?
    \item (How) does the diffusion coefficient depend on electron energy?
    \item What is the maximum energy up to which electrons can be found in the Halo?
\end{itemize}
Answers to these questions will help determine the processes that lead to the suppressed diffusion and the formation of TeV Halos, and let us better constrain the number of electrons from such sources that could make it to Earth.

\subsection{Survey of known pulsars}
A deep survey, looking for extended TeV emission around known radio and gamma-ray pulsars, will provide a larger sample of TeV Halos and enable population studies. The following questions should be addressed:
\begin{itemize}[noitemsep]
    \item How common are TeV Halos?
    \item How and when in a pulsar's lifetime do they form? How do they evolve?
    \item How similar or dissimilar are these objects to one another in terms of the gamma-ray emissivity and the diffusion coefficient?
\end{itemize}
Under the hypothesis that all pulsars have TeV Halos with the same efficiency to produce gamma rays as Geminga, an all-sky VHE gamma-ray survey of with milli-Crab sensitivity should be able to discover several tens, up to hundreds of such objects, depending on how long it takes for a TeV Halo to form \cite{Linden2017}. The study of younger pulsars in the age range of 10 kyr--100 kyr is especially crucial to constrain the minimum age  and formation mechanisms. If no or few new TeV Halos are detected, we will be able to place constraints on the electron acceleration/escape efficiency and the diffusion coefficients around known pulsars.

\subsection{Blind searches for new pulsars}
\renewcommand{\bottomfraction}{0.8}
\begin{figure}[tb]

\includegraphics[width=\linewidth, trim=0.5cm 5.65cm 0.cm 0cm, clip]{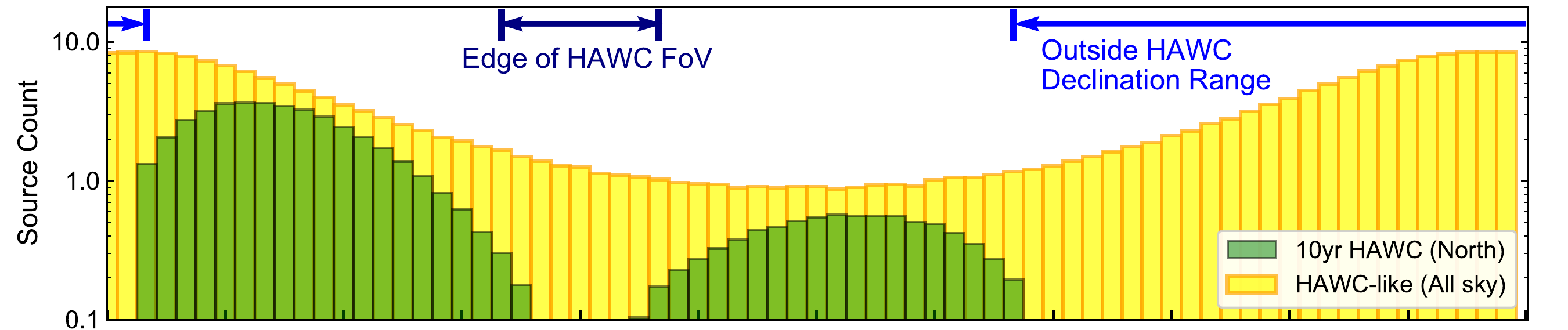}\\ 
\includegraphics[width=\linewidth, trim=0.5cm 0.cm 0.cm 10.4cm, clip]{figures/spatial.pdf} 

  \caption{Prediction for the number and Galactic Longitude distribution of TeV Halos to be discovered with current or future instruments. A ``HAWC-like" 10 year sensitivity is modeled as a theoretical instrument with a sensitivity of 3\% of the Geminga flux across the entire sky. Figure from Ref.\,\cite{Sudoh:2019lav} (see there for more details about the model), reproduced with permission.
  }
  \label{fig:new_halos}
\end{figure}

So far, most pulsars have been detected via searches for periodic radio pulses. Some have also been found via their pulsed gamma-ray emission. However, there is an unknown fraction of `mis-aligned' pulsars, whose radio-beam does not sweep Earth and from which we do not observe any pulsed emission. Blind searches for TeV Halos in the Milky way and neighboring galaxies such as the Large Magellanic Cloud are expected to lead to the discovery and identification of tens or hundreds new pulsars (see \Cref{fig:new_halos} and Ref.\,\cite{Sudoh:2019lav}). In addition to further improving our understanding of TeV Halos and their evolution, blind searches will lead to a better understanding of pulsar populations and improve measurements of (see Ref.\,\cite{Sudoh:2019lav}):
\begin{itemize}[noitemsep]
    \item the total number of pulsars and the pulsar beaming fraction,
    \item the distribution of pulsar parameters such as the initial spin period and the magnetic field strength, and how these distributions depend on the host galaxy, and
    \item the contribution of unresolved TeV Halos to the diffuse gamma-ray background.
\end{itemize} 


\section{Instrumentation}

\subsection{Current and planned instrumentation}
While space-based gamma-ray detectors typically have large fields-of-view (FoVs), very good angular and energy resolution, and low backgrounds, they are limited in detector volume and thus cannot compete with ground-based observatories at TeV energies and above. 

For ground-based gamma-ray detectors, the dominant detection techniques are imaging air-Cherenkov telescope (IACT) arrays, which record stereoscopic images of gamma-ray and cosmic-ray induced air showers via Cherenkov light emitted as the shower propagates through the atmosphere, and particle detector ground arrays, which detect air shower particles as they reach the ground. IACTs typically have smaller instantaneous FoVs (few degrees radius) compared to ground arrays ($\sim 45^\circ$), but better resolution and background suppression. Both techniques are sensitive to gamma-ray showers from tens/hundreds of GeV to hundreds of TeV, depending on the detector altitude. Both are inherently modular: the instruments' sensitivities can be increased by increasing the number of telescopes/detector  units.

The current generation of IACTs (VERITAS \cite{2015ICRC...34..771P}, MAGIC \cite{ALEKSIC201661}, and FACT \cite{Anderhub_2013} in the northern hemisphere and H.E.S.S.\,\cite{Hinton:2004eu} in the south), sensitive in the energy range from tens of GeV to tens of TeV and beyond, will continue operating for the next few years. HAWC \cite{Abeysekara:2017hyn} in the northern hemisphere is the only active ground array gamma-ray detector. It was recently extended by an outrigger array and there are plans to run until at least 2025.

Two instruments are currently in the prototype/construction phase: LHAASO \cite{LHAASO}, a hybrid ground/air Cherenkov array located in the northern hemisphere, will start partial operations this year (full operations in 2020). It is planned to operate for at least 10--15 years. LHAASO will have a wide FoV and be sensitive to gamma rays from 100 GeV to 1 PeV. The Cherenkov Telescope Array CTA \cite{CTA_ScienceTDR} will start partial operations in 2022 
and full operations in 2025. It comprises two sites, one in each hemisphere, covering an energy range from 20 GeV to at least 300 TeV. CTA is an array of pointing instrument with FoVs of several degrees radius and its observing program includes a combination of pointed and survey observations.

\subsection{Recommendations for future projects}
To perform the studies outlined in \cref{sec:studies}, improvements in VHE gamma-ray detection are necessary. We need both sensitive pointing instruments with good angular resolution for precise measurements of the TeV Halo morphology, as well as surveys covering the Galactic plane and nearby galaxies to discover more TeV Halos and characterize their populations. Due to the distributions of pulsars in the Galaxy, more Galactic TeV Halos are expected to be found closer to the Galactic center, visible in the southern sky (see \Cref{fig:new_halos}).

CTA will provide pointed observations in both hemispheres, and LHAASO will survey the northern sky. A sensitive TeV gamma-ray survey observatory located in the southern hemisphere, with a good view of the inner Galaxy, is needed to maximize the scientific return of searches for new TeV Halos. The planned TeV instruments would thus be well complemented by a next-generation TeV-range gamma-ray survey instrument located in the southern hemisphere, such as the proposed SGSO project \cite{SGSOwhitepaper}.

\clearpage
\printbibliography

\end{document}